\documentclass[superscriptadress,citeauthorscript,aip,preprintnumbers,amsmath,amssymb,reprint,graphicx,amsmath,amssymb,twocolumn]{revtex4-2}
\usepackage{graphicx}
\usepackage{dcolumn}
\usepackage{bm}
\usepackage{color}
\usepackage{gensymb}
\usepackage[utf8]{inputenc}
\usepackage[T1]{fontenc}
\usepackage{mathptmx}
\usepackage{etoolbox}
\usepackage{url}
\usepackage{hyperref}
\usepackage[caption=false]{subfig}

\makeatletter
\def\@email#1#2{%
 \endgroup
 \patchcmd{\titleblock@produce}
  {\frontmatter@RRAPformat}
  {\frontmatter@RRAPformat{\produce@RRAP{*#1\href{mailto:#2}{#2}}}\frontmatter@RRAPformat}
  {}{}
}%
\makeatother
\begin{document}

\preprint{LA-UR-26-29334}

\title{Magnetic Johnson noise from arbitrarily shaped conductors: experimental validation of the fluctuation–dissipation finite-element method}


\author{S.~M.~Clayton}
\email{sclayton@lanl.gov}
\affiliation{Los Alamos National Laboratory, Los Alamos, New Mexico 87545, USA}

\author{Y. J. Kim}
\email{youngjin@lanl.gov}
\affiliation{Los Alamos National Laboratory, Los Alamos, New Mexico 87545, USA}

\author{T. E. Chupp}
\affiliation{Department of Physics, University of Michigan, Ann Arbor, Michigan 48109, USA}

\author{A.~Epstein}
\affiliation{Department of Physics, Arizona State University, Tempe, Arizona 85281, USA}

\author{J.~M.~Feltman}
\affiliation{Department of Physics, Arizona State University, Tempe, Arizona 85281, USA}

\author{A.~Fratangelo}
\affiliation{Los Alamos National Laboratory, Los Alamos, New Mexico 87545, USA}

\author{T.~M.~Ito}
\email[Corresponding author: ]{ito@lanl.gov}
\affiliation{Los Alamos National Laboratory, Los Alamos, New Mexico 87545, USA}


\author{I.~J.~Rhudy}
\affiliation{Department of Physics, University of Michigan, Ann Arbor, Michigan 48109, USA}

\author{F.~H.~Sottrel}
\affiliation{Department of Physics, University of Michigan, Ann Arbor, Michigan 48109, USA}

\author{W.~A.~Terrano}
\affiliation{Department of Physics, Arizona State University, Tempe, Arizona 85281, USA}

\author{K.~L.~Wood}
\affiliation{Department of Physics, Arizona State University, Tempe, Arizona 85281, USA}

\date{\today}

\begin{abstract}
Magnetic Johnson noise--magnetic noise caused by thermal current fluctuations in conducting materials--can interfere with measurements and limit their sensitivity. As the sensitivity of experiments increases, it becomes increasingly important to accurately estimate the magnetic Johnson noise from experimental components with complex shapes during both the design and analysis phases. A particularly powerful approach is to combine the fluctuation-dissipation theorem with finite-element calculations.  This versatile approach can be used to calculate the magnetic Johnson noise from conductors of arbitrary geometry for a wide range of frequencies and for magnetic field detectors with general geometries, such as a finite-sized loop or a gradiometer. Using a SQUID gradiometer array, we measured the magnetic Johnson noise from conductors with complex geometries and compared the results with calculations performed using this method. The calculations reproduce both the measured noise amplitude and its frequency dependence, providing an experimental validation of this method. We also performed a detailed analysis of how the geometry and material parameters affect the frequency dependence of magnetic Johnson noise. 
\end{abstract}

\maketitle

\section{\label{sec:introduction}Introduction}
Thermal current fluctuations in conducting materials generate fluctuating magnetic fields that can limit the sensitivity of precision measurements. This magnetic Johnson noise has been investigated in a variety of experimental settings, including biomagnetic measurements based on superconducting quantum interference device (SQUID) sensors,\cite{Varpula1984,Nenonen1996} applications involving trapped atoms,\cite{Henkel2005} searches for permanent electric dipole moments (EDMs) of atoms and neutrons,\cite{Lamoreaux1999,Munger2005,Lee2008,Ayres2021,Blatnik2026}
and high-frequency magnetic measurements using conductive radio-frequency shields.\cite{Savukov2020}


Estimates of magnetic Johnson noise are typically performed using one of two methods: the direct method or the reciprocal method. In the direct method, the fluctuating current density is given by the conductivity of the material through the Johnson-Nyquist theorem,\cite{Johnson1928,Nyquist1928} and the resulting fluctuating  magnetic field is calculated either by solving Maxwell's equations\cite{Varpula1984,Nenonen1996} or by applying the Biot-Savart law.\cite{Lamoreaux1999,Sandin2010,Ready2021} The reciprocal method, on the other hand, makes use of the fluctuation-dissipation (F-D) theorem,\cite{Callen1951,Kubo1966} of which the Johnson-Nyquist theorem is a special case. With this method, the magnetic Johnson noise (fluctuation) is determined from the power dissipated in the conductor body (dissipation) when it is driven by an oscillating magnetic field.

Both methods have been used to derive analytic formulas or general expressions for some specific geometries.\cite{Varpula1984,Nenonen1996,Kasai1993,Roth1998,Clem1987,Sidles2003,Lee2008} 
A numerical implementation of the direct approach for arbitrarily shaped thin conductors, including frequency dependence and spatial correlations, has also been developed.\cite{Iivanainen2021}
However, calculating the magnetic Johnson noise using the direct method for general three-dimensional conductors remains a daunting task. As the sensitivity of experiments increases, it becomes more important to accurately estimate the possible magnetic Johnson noise from various experimental components during the design phase (see e.g. Ref.~\onlinecite{Blatnik2026}). Such an  ability is also important for understanding the noise observed by sensitive magnetic sensors, such as SQUIDs.  

Numerical implementations of the reciprocal approach have also been used to calculate magnetic noise in electron-optical systems\cite{Uhlemann2015} and, using finite-element calculations, in magnetic shielding and conducting samples relevant to biomagnetic measurements.\cite{Storm2017,Storm2019}
In our previous work,\cite{Phan2024} we pointed out that the reciprocal method, combined with a finite element method (FEM) tool such as COMSOL\textsuperscript{\textregistered},\cite{COMSOL} provides a practical way to calculate magnetic Johnson noise for arbitrary geometries and frequencies. We demonstrated this method by comparing results obtained using the method to those obtained from analytical formulas or numerical calculations for several specific cases. We showed furthermore that this method can be used, not only to calculate magnetic Johnson noise evaluated at a point in space, but also to calculate magnetic Johnson noise detected by a magnetic field detector with a more general shape, such as a finite-sized loop, a gradiometer, etc., taking into account the correlations of magnetic fields detected by different parts of such a detector.

More concretely, when evaluating magnetic Johnson noise at a given position near a conductor, a hypothetical current loop of area $A$, carrying an oscillating current $I(t)=I_0\sin\omega t$, is placed at that position. The loop geometry corresponds to that of the magnetic field detector: a small loop for evaluating the noise at a point in space, a finite-sized loop for evaluating the noise detected by such a loop, or two oppositely wound loops for evaluating the noise detected by a gradiometer, etc. The resulting oscillating magnetic field induces eddy currents in the conductor, producing a time-averaged power dissipation $P(f)$. The F-D theorem relates the one-sided amplitude spectral density of magnetic Johnson noise, $B_n(f)$, to this dissipated power:
\begin{equation}
B_n(f) = \frac{\sqrt{4kT}\sqrt{2P(f)}}{2\pi fANI_0},
\label{eq:F-Dtheorem}
\end{equation}
where $T$ is the temperature of the conductor, $k$ is the Boltzmann constant, $N$ is the number of turns in the current loop, and $\omega=2\pi f$ is the angular frequency of the driving current. Finite-element analysis tools such as COMSOL make it straightforward to compute the time-averaged power dissipated in the conductor for a specified current-loop geometry and driving frequency.

To provide an experimental demonstration of this F-D FEM method, we used a SQUID gradiometer array to measure the magnetic Johnson noise from conductors with geometries for which it is difficult to calculate the magnetic Johnson noise using analytic formulas. The measured magnetic Johnson noise was compared to calculations performed using the F-D FEM method. The purpose of this paper is to report on these measurements and their comparison to calculations using the F-D FEM method.

The remainder of this paper is structured as follows. In Sections II and III, the experimental setup and measurements are described, respectively. In Section IV, we compare the measured results with calculations based on the F-D FEM method and discuss the results. In Section V, we present our conclusions.

\section{\label{sec:setup}Experimental setup}
The reported measurements were performed as part of the measurements conducted to characterize the SQUID-based detector system for a new $^{129}$Xe EDM search experiment being prepared at Los Alamos National Laboratory.\cite{Kim2026} Details of the apparatus used, including the SQUID detector system, are described in Ref.~\onlinecite{Kim2026}. Here we give a brief account relevant to the content of this paper. 

The measurements were performed inside a magnetically shielded room (MSR) with an internal dimension of 2.9~m (width) $\times$ 3.8~m (depth) $\times$ 2.4~m (height). The walls of the MSR were made of two layers of high permeability alloy, providing a residual magnetic field of 2~nT.\cite{Kim2015}

\begin{figure}
    \centering
    \includegraphics[width=1.0\linewidth]{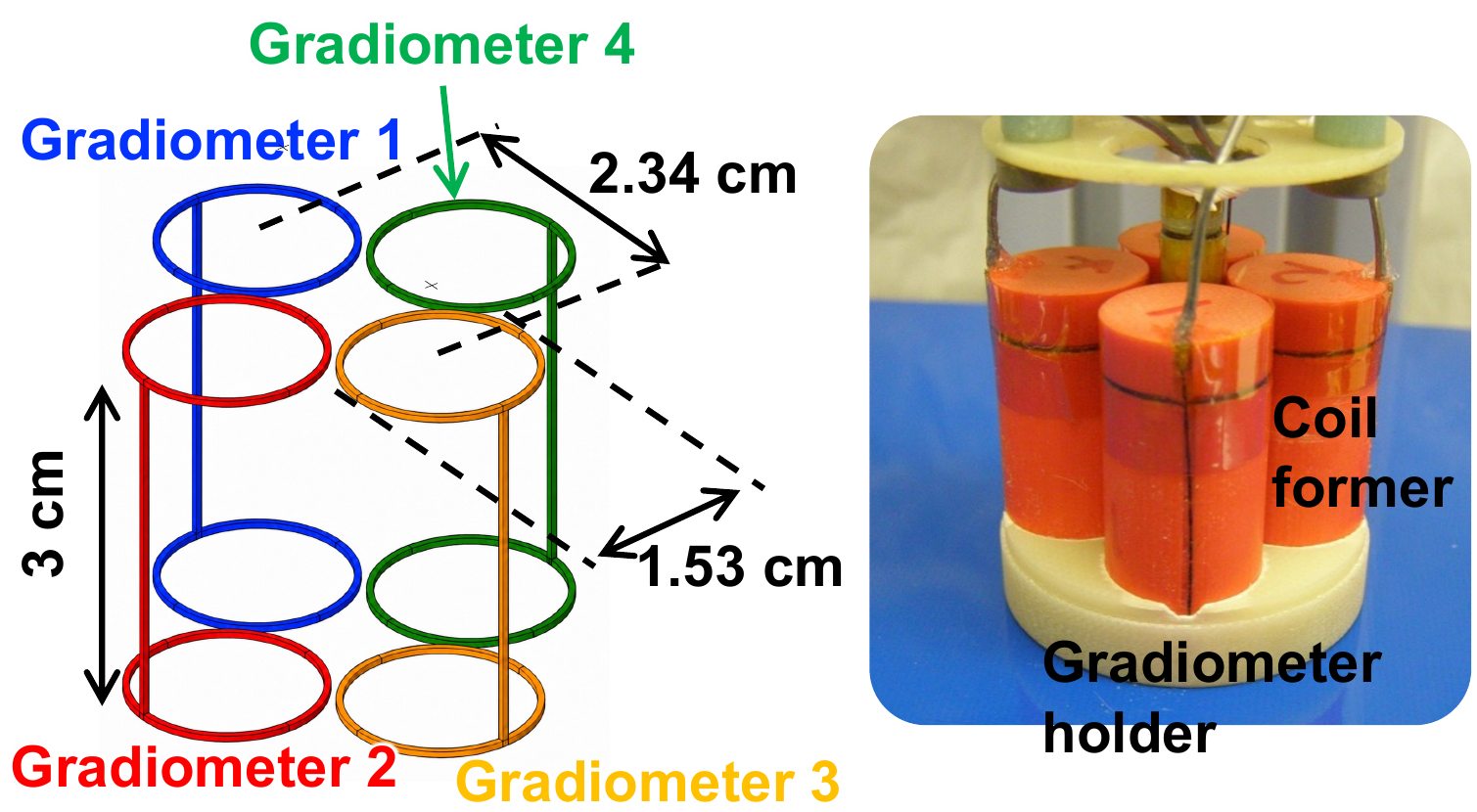}
    \caption{A schematic (left) and a photograph (right) of the gradiometer array.}
    \label{fig:gradiometer_array}
\end{figure}

An array of SQUID-based gradiometers was used as the magnetic field sensor. Four two-turn axial gradiometers were fabricated by manually winding 150-$\mu$m niobium-titanium (NbTi) wire onto fiberglass coil forms and mounting them on the same rigid fiberglass platform (Fig.~\ref{fig:gradiometer_array}). They were connected via 10-cm-long twisted-pair NbTi leads to STAR Cryoelectronics\cite{StarCryo} SQ180 SQUIDs. The SQUIDs were shielded in a lead enclosure, which provided radio-frequency (RF) shielding. Furthermore, the NbTi lead wires were also enclosed in lead shielding to suppress noise pickup. 

The SQUID–gradiometer system was housed in a 4-liter low-noise fiberglass liquid-helium Dewar (Tristan Technologies\cite{Tristan} BMD-4), with a warm-cold distance of 0.86~cm. The signals from the SQUID-gradiometer system were digitized using a D-TACQ\cite{D-TACQ} ACQ437ELF DAQ system at a sampling rate of 43.5~kHz. A 4-pole Butterworth low-pass filter with a cutoff frequency of $f_c \sim 3$~kHz was used in the SQUID signal chain. 

\begin{figure}
    \centering
    \includegraphics[width=1\linewidth]{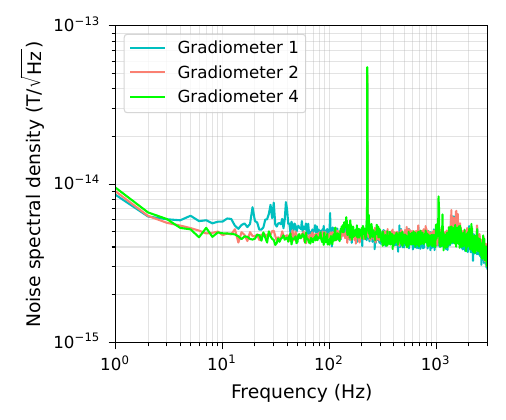}
    \caption{The baseline noise spectrum of the SQUID gradiometers in the absence of any samples.}
    \label{fig:noise_floor}
\end{figure}

The measured noise level without any samples is shown in Fig.~\ref{fig:noise_floor}. The noise floor was $\lesssim $5~fT/$\sqrt{\rm Hz}$ for the frequency range of $\sim$5~Hz to 3~kHz. We observe several peaks, with the most prominent one at $\sim$226~Hz. These are attributed to noise from nearby instruments. Note that for the measurement reported here, gradiometer \#3 was not operational. 

The measurement was performed at room temperature. However, the temperature of the sample was not measured. For all the calculations shown in Sec.~\ref{sec:comparison}, a temperature of 300~K was assumed. An uncertainty of $\pm10$~K is assigned to the temperature. The placement of the samples is discussed in the next section. The distance between the bottom of the Dewar tail and each sample was measured to $\pm 1$~mm. Considering the possible tilt of the sample with respect to the gradiometer loop assembly, we assign an uncertainty of $\pm 2$~mm to the distance between the sample and each gradiometer. Since the sample-gradiometer distance is the dominant geometric uncertainty, we did not consider uncertainties in the directions perpendicular to the gradiometer axis. The systematic effects associated with temperature and geometry uncertainties are also discussed in Sec.~\ref{sec:comparison}.

\section{\label{sec:measurements}Measurements}
\begin{figure}
    \centering
    \includegraphics[width=0.5\linewidth]{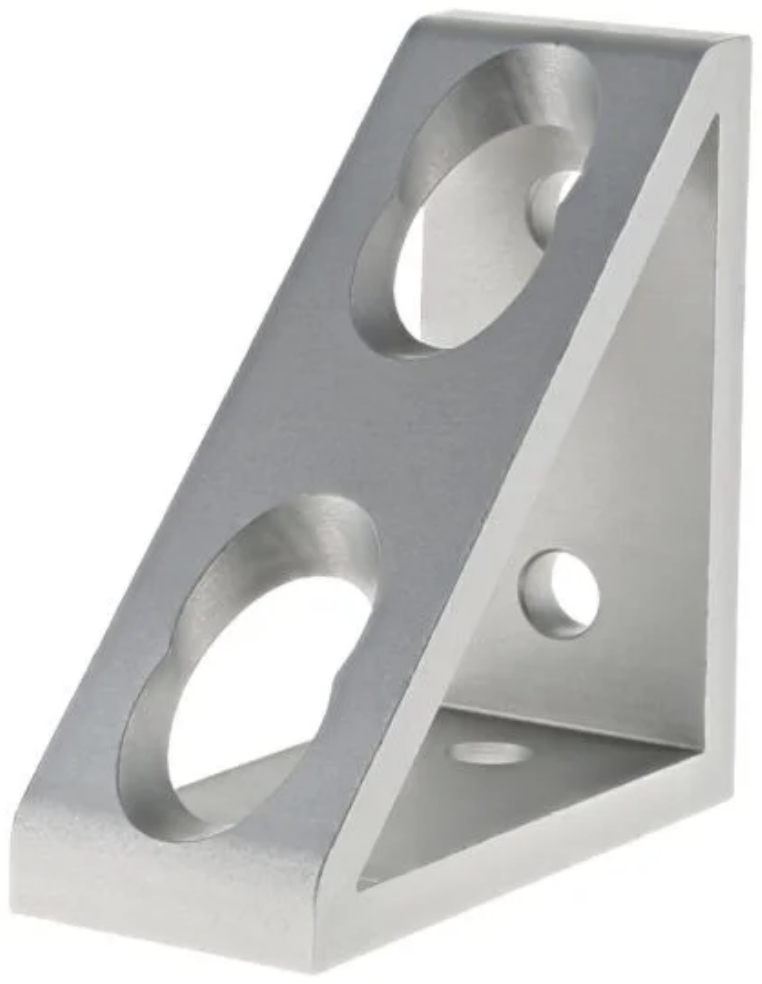}
    \caption{A photograph of the corner bracket for T-slot structural framing used in this study. Taken from the manufacturer's website.\cite{80-20}}
    \label{fig:80-20photo}
\end{figure}

Four different geometries were studied. One employed a plate made of aluminum alloy Al6061-T6 with dimensions of 1.9~cm $\times$ 10.2~cm $\times$ 30.0~cm as the conductor. The remaining three employed a corner bracket for the 80/20\textsuperscript{\textregistered} T-slot structural framing system\cite{80-20} (see Fig.~\ref{fig:80-20photo}). The material of the 80/20 bracket was aluminum alloy Al6063-T6. In each measurement, the data were recorded for $100-200$~s. Below we describe each geometry as well as the magnetic Johnson noise results.

\subsection{Magnetic Johnson noise from an Al6061-T6 alloy plate}
\begin{figure}
    \centering
    \includegraphics[width=1.0\linewidth]{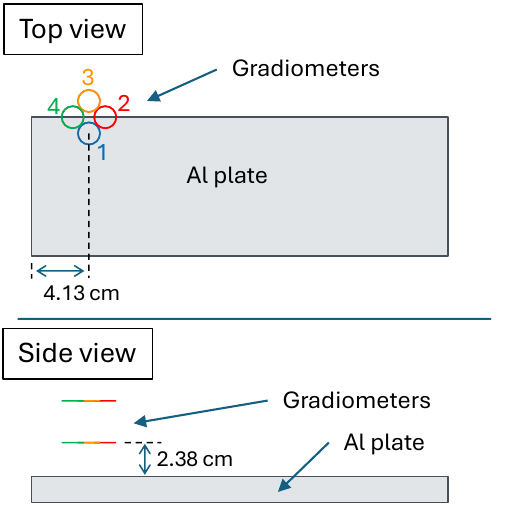}
    \caption{A schematic showing the geometry in which the magnetic Johnson noise from the Al6061-T6 plate was measured. }
    \label{fig:Al_plate_geom}
\end{figure}

\begin{figure}
    \centering
    \includegraphics[width=1.0\linewidth]{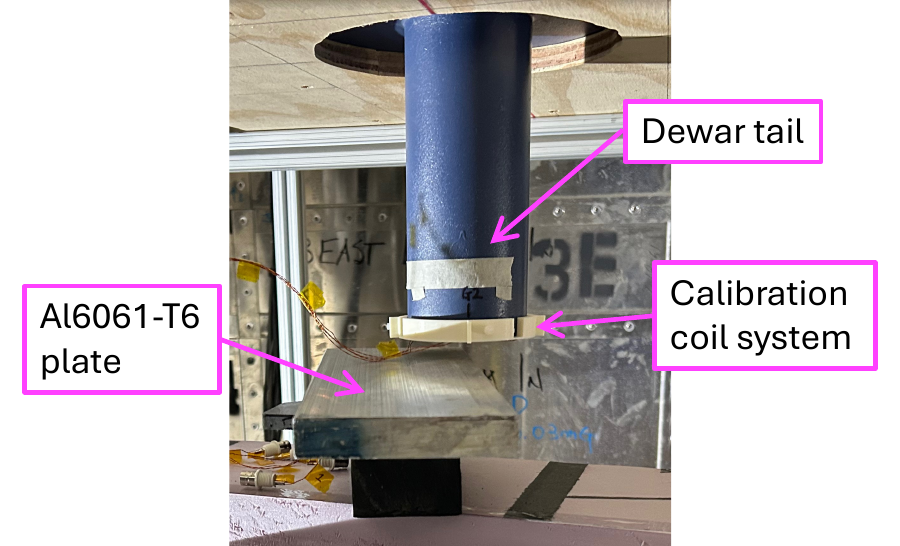}
    \caption{A photograph of the setup with the Al6061-T6 plate. The calibration coil system, composed of four wire loops mounted on a rotatable plate, allowed us to determine the azimuthal orientation of the gradiometer system.}
    \label{fig:Al_plate_photo}
\end{figure}

\begin{figure}
    \centering
    \includegraphics[width=1\linewidth]{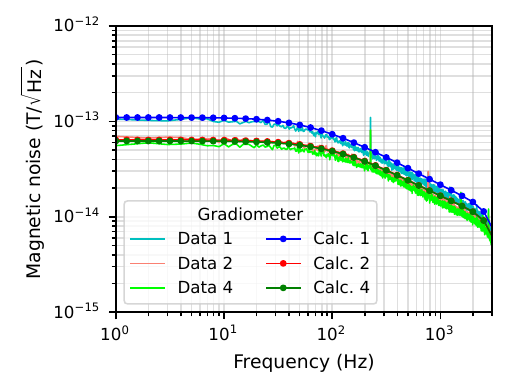}
    \includegraphics[width=1\linewidth]{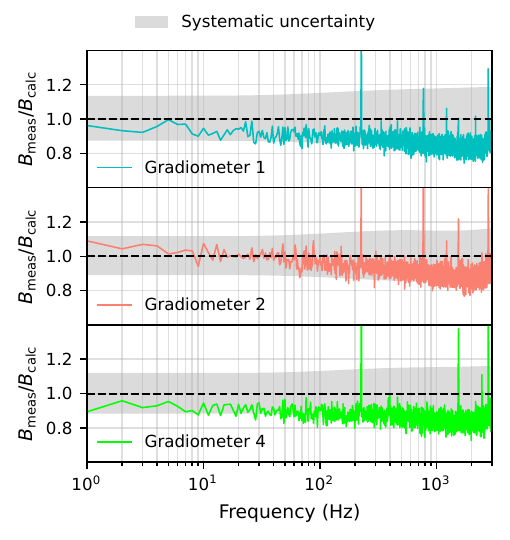}
    \caption{Top panel: Measured magnetic Johnson noise from the Al6061-T6 plate along with calculated values obtained using the F-D FEM method. Bottom panel: Ratio of the measured magnetic Johnson noise to the calculated value for each gradiometer. The dashed horizontal line indicates perfect agreement. The gray band represents the combined uncertainty from the gradiometer–sample distance, electrical conductivity, and mesh discretization.}
    \label{fig:Al_plate}
\end{figure}

We placed the Al6061-T6 plate in the position shown in  Fig.~\ref{fig:Al_plate_geom}. A photograph of the setup is shown in Fig.~\ref{fig:Al_plate_photo}. We chose this arrangement so that the gradiometer array was near one of the edges to test the method in a geometry with substantial edge effects. The measured magnetic noise is shown in the top panel of Fig.~\ref{fig:Al_plate}.

\subsection{Magnetic Johnson noise from a corner bracket for T-slot structural framing}
To demonstrate the ability to calculate the magnetic Johnson noise for complex geometries, we measured the magnetic Johnson noise from an 80/20 corner bracket that we found in our laboratory (see Fig.~\ref{fig:80-20photo}). 

\begin{figure}
    \centering
    \includegraphics[width=1.0\linewidth]{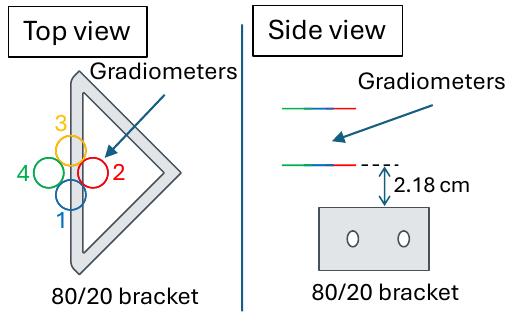}
    \caption{A schematic showing Geometry 1 with the 80/20 corner bracket.}
    \label{fig:80-20_geom_1_geom}
\end{figure}

\begin{figure}
    \centering
    \includegraphics[width=1.0\linewidth]{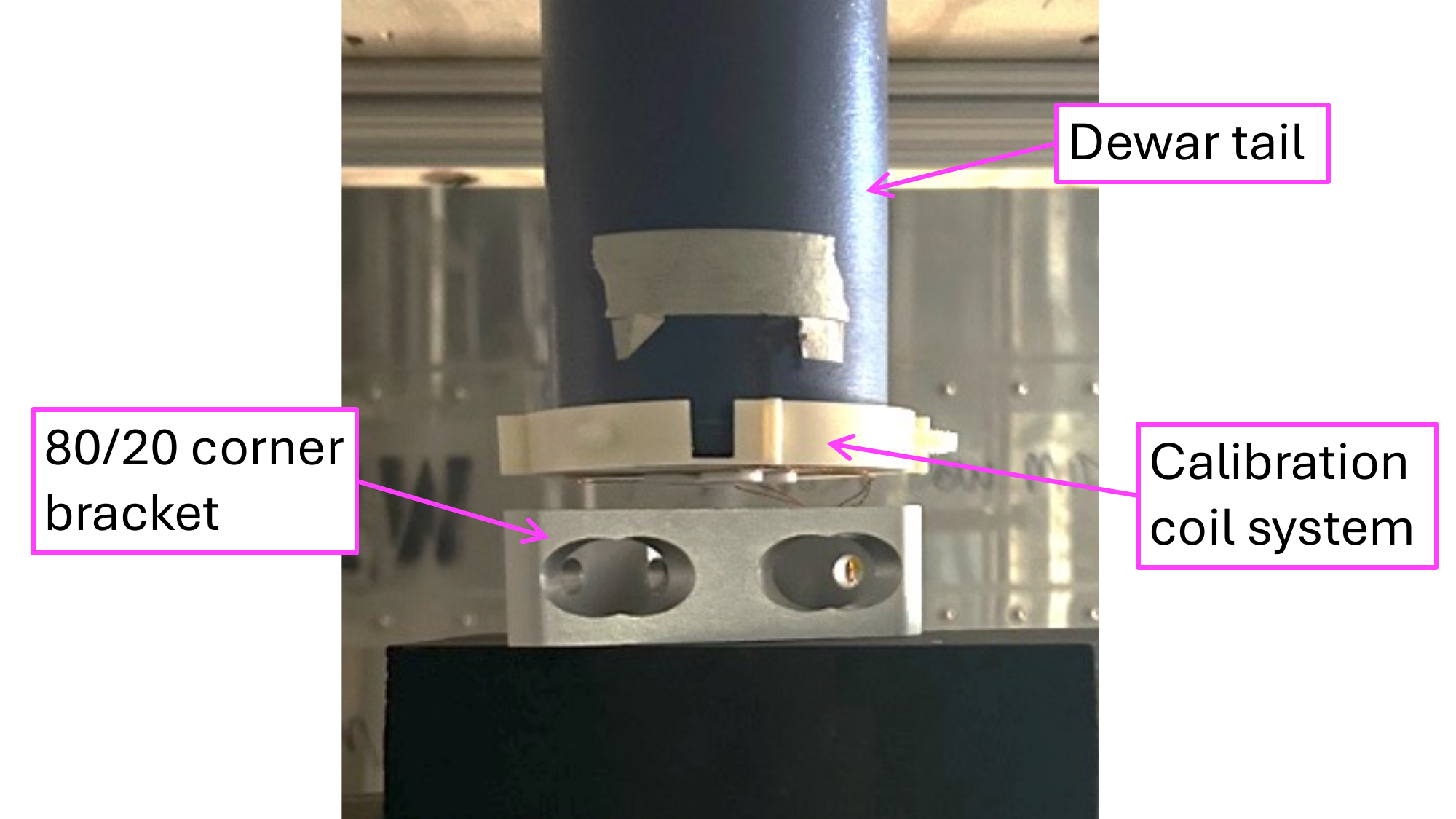}
    \caption{A photograph of the setup with the 80/20 corner bracket in Geometry 1. The calibration coil system, composed of four wire loops mounted on a rotatable plate, allowed us to determine the azimuthal orientation of the gradiometer system.}
    \label{fig:Al_80-20_photo}
\end{figure}

\begin{figure}
    \centering
    \includegraphics[width=1.0\linewidth]{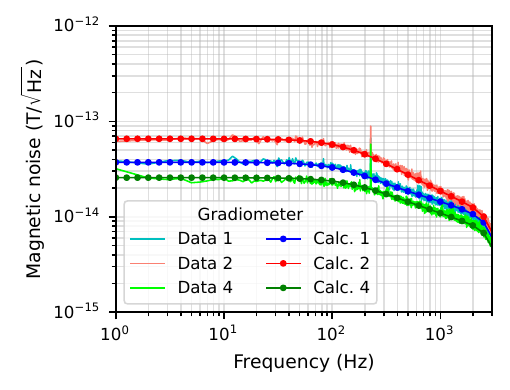}
        \includegraphics[width=1\linewidth]{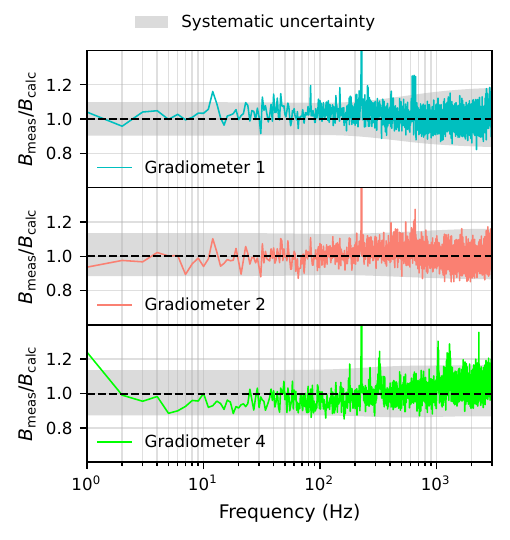}
    \caption{Top panel: Measured magnetic Johnson noise from the 80/20 corner bracket in Geometry 1 along with calculated values obtained using the F-D FEM method. Bottom panel: Ratio of the measured magnetic Johnson noise to the calculated value for each gradiometer. The dashed horizontal line indicates perfect agreement. The gray band represents the combined uncertainty from the gradiometer–sample distance, electrical conductivity, and mesh discretization.}
    \label{fig:80-20_geom_1}
\end{figure}

\begin{figure}
    \centering
    \includegraphics[width=1\linewidth]{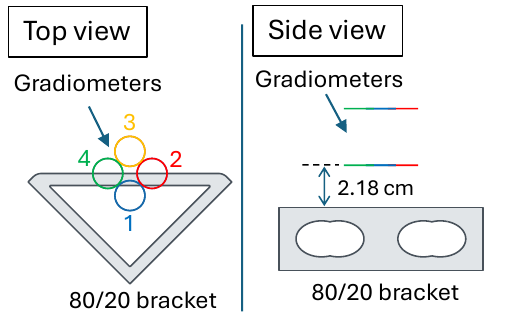}
    \caption{A schematic showing Geometry 2 with the 80/20 corner bracket.}
    \label{fig:80-20_geom_2_geom}
\end{figure}

\begin{figure}
    \centering
    \includegraphics[width=1.0\linewidth]{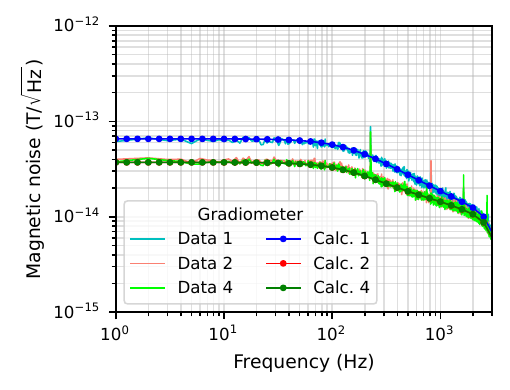}
    \includegraphics[width=1\linewidth]{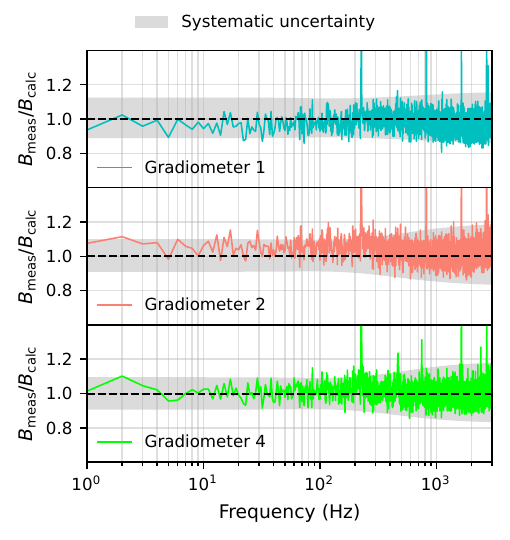}
    \caption{Top panel: Measured magnetic Johnson noise from the 80/20 corner bracket in Geometry 2 along with calculated values obtained using the F-D FEM method. Bottom panel: Ratio of the measured magnetic Johnson noise to the calculated value for each gradiometer. The dashed horizontal line indicates perfect agreement. The gray band represents the combined uncertainty from the gradiometer–sample distance, electrical conductivity, and mesh discretization.}
    \label{fig:80-20_geom_2}
\end{figure}

\begin{figure}
    \centering
    \includegraphics[width=1\linewidth]{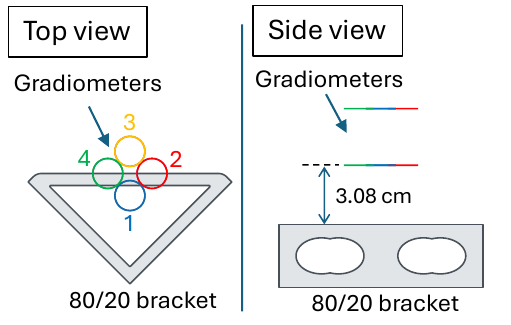}
    \caption{A schematic showing Geometry 3 with the 80/20 corner bracket.}
    \label{fig:80-20_geom_3_geom}
\end{figure}

\begin{figure}
    \centering
    \includegraphics[width=1.0\linewidth]{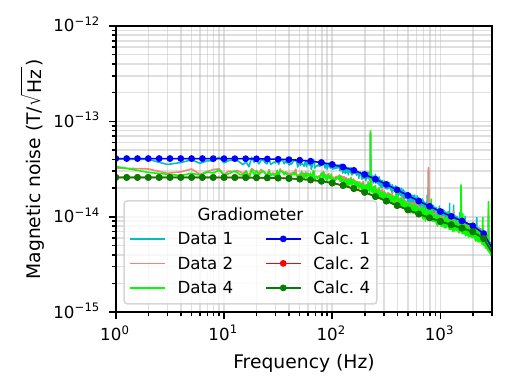}
    \includegraphics[width=1\linewidth]{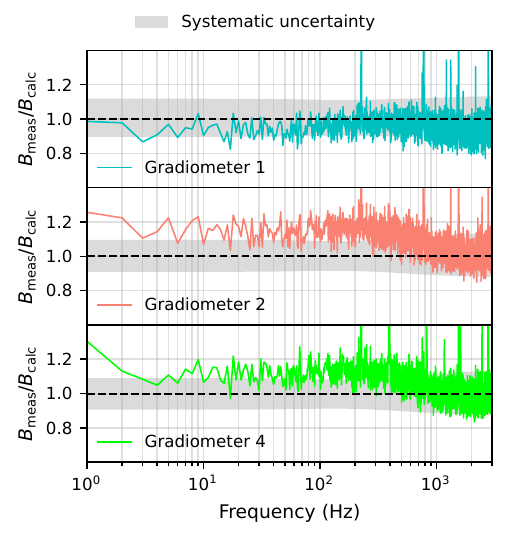}
    \caption{Top panel: Measured magnetic Johnson noise from the 80/20 corner bracket in Geometry 3 along with calculated values obtained using the F-D FEM method. Bottom panel: Ratio of the measured magnetic Johnson noise to the calculated value for each gradiometer. The dashed horizontal line indicates perfect agreement. The gray band represents the combined uncertainty from the gradiometer–sample distance, electrical conductivity, and mesh discretization.}
    \label{fig:80-20_geom_3}
\end{figure}

We used three different geometries. Geometry 1 is shown schematically in Fig.~\ref{fig:80-20_geom_1_geom} and photographically in Fig.~\ref{fig:Al_80-20_photo}, with the measured magnetic Johnson noise plotted in the top panel of Fig.~\ref{fig:80-20_geom_1}. Geometry 2 was obtained by rotating the bracket in Geometry 1 by 90 degrees clockwise; its schematic and measured noise are shown in Fig.~\ref{fig:80-20_geom_2_geom} and the top panel of Fig.~\ref{fig:80-20_geom_2}, respectively. In Geometry 3, the bracket is lowered by 9~mm relative to its position in Geometry 2; the corresponding schematic and measured noise are shown in Fig.~\ref{fig:80-20_geom_3_geom} and Fig.~\ref{fig:80-20_geom_3}.

\subsection{Magnetic Johnson noise from a corner bracket for T-slot structural framing detected by a second-order gradiometer }

\begin{figure}
    \centering
    \includegraphics[width=1\linewidth]{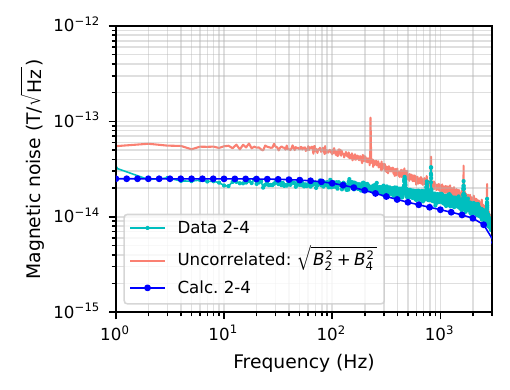}
    \caption{Magnetic noise measured by the combined gradiometer formed from gradiometers 2 and 4 in Geometry 2, compared with the F-D FEM calculation. Also shown is the quadrature sum of the individual measured noise amplitudes, representing the expected difference-signal noise in the absence of correlation.}
    \label{fig:4th_order_grad}
\end{figure}

In addition, we formed a second-order gradiometer from gradiometers 2 and 4 in the offline analysis of the acquired data. The magnetic Johnson noise detected by such a second-order gradiometer for Geometry 2 with the 80/20 corner bracket is shown in Fig.~\ref{fig:4th_order_grad}. This tests whether the calculation reproduces spatial correlations, beyond matching the individual noise spectra.

\section{\label{sec:comparison}Comparison with calculations and discussion}
The magnetic Johnson noise detected by the gradiometers was calculated using the F-D FEM method as described in Sec.~\ref{sec:introduction} (for more details, see Ref.~\onlinecite{Phan2024}). COMSOL was used to calculate the time-averaged power dissipated in the sample by the oscillating magnetic field, which in turn was created by an oscillating current flowing in the wire that forms the gradiometer loops. The input to the COMSOL model was the geometry, that is the geometry of the gradiometer loops as well as the shape and location of the sample (Al6061-T6 plate and the 80/20 corner bracket), and the electrical conductivity of the material of the sample. For the shape of the 80/20 corner bracket, a CAD file was downloaded from the manufacturer's website\cite{80-20} and was imported into the COMSOL model. 

\begin{figure}
    \centering
    \includegraphics[width=1.0\linewidth]{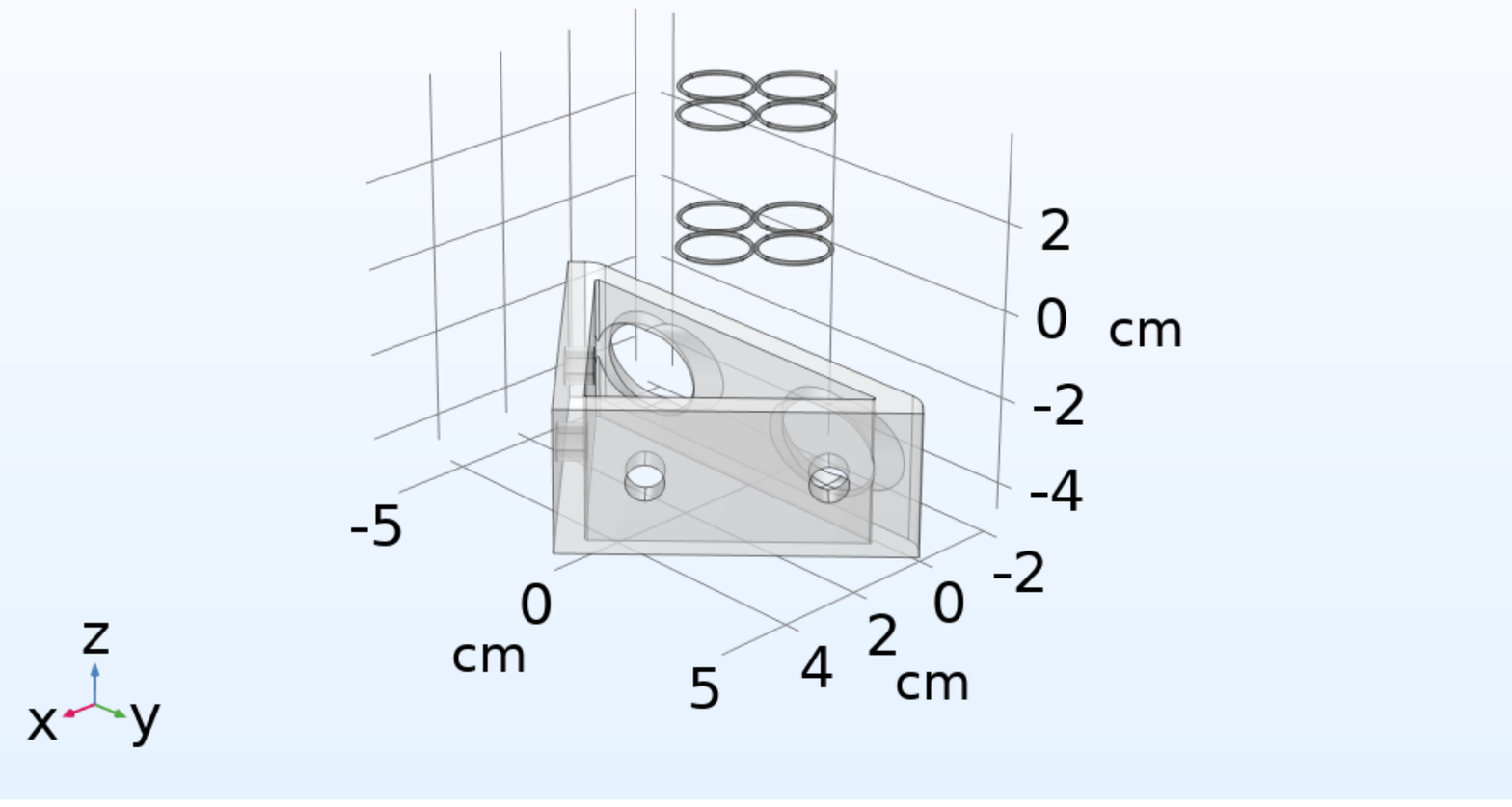}
    \caption{Geometry used in the COMSOL calculation.}
    \label{fig:80-20_comsol}
\end{figure}

The geometry used in the COMSOL calculation with the 80/20 corner bracket is shown in Fig.~\ref{fig:80-20_comsol}.
Although all four gradiometers are shown in the figure to indicate their relative positions, the calculation was performed separately for each operational gradiometer. In each calculation, only the current loops representing that gradiometer were driven, with the currents in the two loops flowing in opposite directions. The resulting total time-averaged power dissipated in the conductor was then used in 
Eq.~(\ref{eq:F-Dtheorem}) to calculate the magnetic Johnson noise detected by that gradiometer.

For the second-order gradiometer formed from gradiometers 2 and 4, the loops representing both gradiometers were driven simultaneously, with the relative current polarity chosen to represent the difference of their measured signals. The resulting total time-averaged power dissipated in the sample was used in Eq.~(\ref{eq:F-Dtheorem}) to calculate the noise in this difference signal, thereby accounting for the spatial correlations between the fields detected by the two gradiometers.

For the electrical conductivity, $\sigma = 2.46\times 10^7$~S/m, a typical value for Al6061-T6 was used for the Al6061-T6 plate and $\sigma = 3.030\times 10^7$~S/m, a typical value for Al6063-T6 was used for the 80/20 corner bracket. Note that there are $\pm 8$\% and $\pm 9$\% variations in the exact conductivity\cite{USAF2014} for Al6061-T6 and Al6063-T6 respectively due to allowed variations in the alloy composition and variability in the tempering process.

Baseline noise consistent with Fig.~\ref{fig:noise_floor} was added in quadrature to the obtained magnetic Johnson noise. The resulting total noise was further modified by the function to account for the 4-pole Butterworth low-pass filter:
\begin{equation}
    B_{\rm out}(f) = \frac{B_{\rm in}(f)}{\sqrt{1+\left ( \frac{f}{f_c} \right )^8}},
\end{equation}
where $f_c = 2970$~Hz, determined by fitting the function to the noise spectrum obtained without the object of interest. 

The calculated magnetic Johnson-noise spectra are plotted together with the measured spectra in the top panels of Figs.~\ref{fig:Al_plate}, \ref{fig:80-20_geom_1}, \ref{fig:80-20_geom_2}, and \ref{fig:80-20_geom_3}. The corresponding ratios of the measured to calculated values are shown in the bottom panels. Plots in these figures are cut off at 3~kHz. This is because beyond this frequency the measured noise is dominated by the effect of the low-pass filter. 

\begin{figure}
    \centering
    \includegraphics[width=1\linewidth]{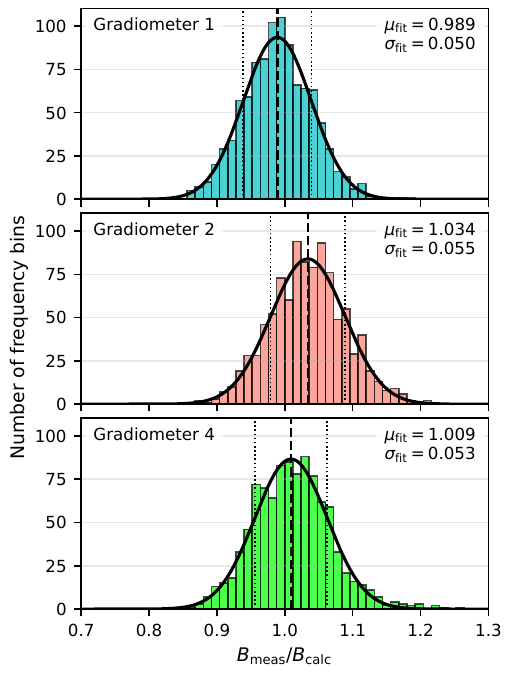}
    \caption{Distribution of the ratio $B_{\rm meas}/B_{\rm calc}$ for 1 to 1000~Hz for Geometry 2 with the 80/20 corner bracket. The solid lines are results of a Gaussian fit.}
    \label{fig:ratio_histograms}
\end{figure}

The point-to-point scatter in the measured-to-calculated ratios reflects the statistical uncertainty of each frequency bin of the measured noise spectra. The statistical uncertainty can be evaluated more quantitatively. The data-taking duration ranged from approximately 100 to 200 s. Each record was divided into nonoverlapping 1-s intervals, corresponding to a frequency-bin width of 1~Hz, and the resulting power spectra were averaged. Depending on the duration of the measurement, each spectrum therefore contained approximately 100–200 independent averages. The fractional statistical uncertainty in the amplitude spectral density, \(1/(2\sqrt{N_{\rm av}})\), where $N_{\rm av}$ is the number of independent averages, ranged from approximately \(3.5\%\) to \(5.0\%\). This is also consistent with the distribution of the ratio of $B_{\rm meas}/B_{\rm calc}$ shown in Fig.~\ref{fig:ratio_histograms} for Geometry 2 with the 80/20 corner bracket. 

The total systematic uncertainty in the ratio is plotted for each gradiometer in the bottom panel of Figs.~\ref{fig:Al_plate}, \ref{fig:80-20_geom_1}, \ref{fig:80-20_geom_2}, and \ref{fig:80-20_geom_3}. The dominant experimental systematic uncertainties come from the conductivity of the sample materials and the geometry. In addition, the bands in the bottom panel of these figures include the calculation error due to COMSOL mesh convergence. The systematic uncertainties in the ratio due to these three sources were added in quadrature. 

\begin{figure}[!t]
    \centering
    \includegraphics[width=1\linewidth]{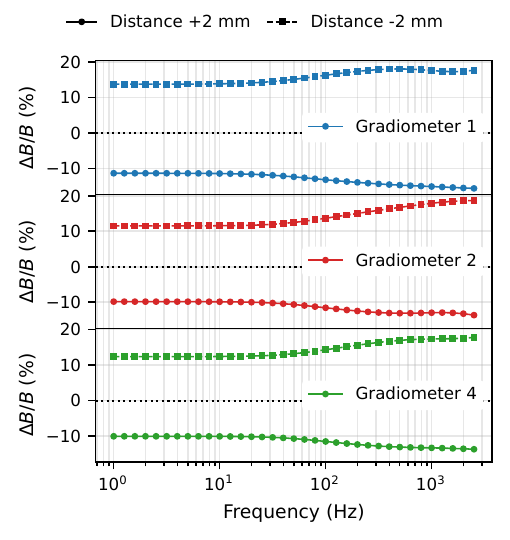}
    \includegraphics[width=1\linewidth]{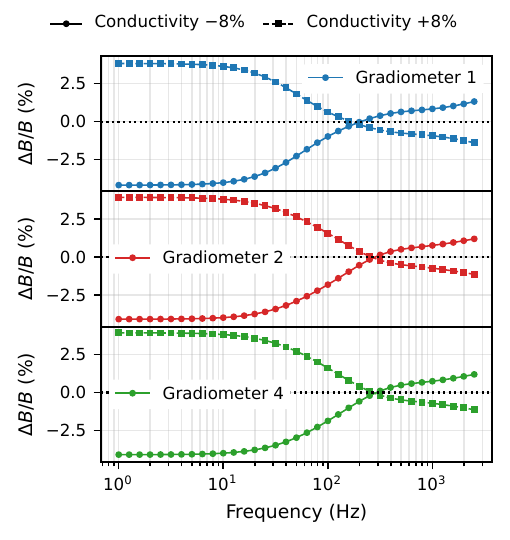}
    \caption{Systematic shift in the calculated magnetic Johnson noise for the Al6061-T6 plate. Top panel: shift in the calculated magnetic noise when the gradiometer--sample distance was shifted by $\pm 2$~mm. Bottom panel: shift in the calculated magnetic Johnson noise when the electrical conductivity of Al6061-T6 was changed by $\pm 8$\%. }
    \label{fig:Al-plate_systematics}
\end{figure}

\begin{figure}[!t]
    \centering
    \includegraphics[width=1\linewidth]{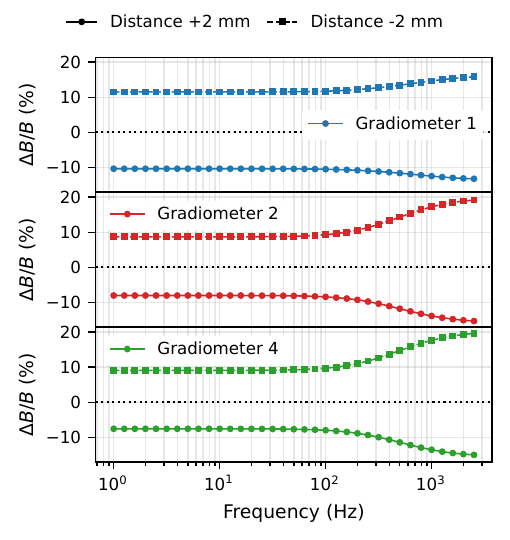}
    \includegraphics[width=1\linewidth]{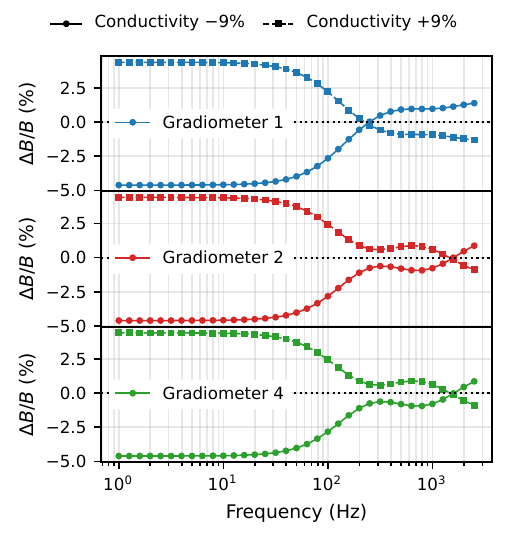}
    \caption{Systematic shift in the calculated magnetic Johnson noise for the 80/20 corner bracket placed in Geometry 2. Top panel: shift in the calculated magnetic noise when the gradiometer--sample distance was shifted by $\pm 2$~mm. Bottom panel: shift in the calculated magnetic Johnson noise when the electrical conductivity of Al6063-T6 was changed by $\pm 9$\%. }
    \label{fig:80-20-geom_2_systematics}
\end{figure}

The distance between the sample and each gradiometer has an uncertainty of up to 2~mm as discussed earlier. The effect of this uncertainty was evaluated directly by repeating the COMSOL calculation after shifting the gradiometer--sample distance by $\pm 2$~mm. We shifted the entire sample vertically by $\pm 2$~mm in the COMSOL model. 
This was used to estimate the effect of the distance uncertainty, including that associated with possible sample tilt.
The resulting change in the calculated magnetic-noise amplitude is shown in the top panel of Figs.~\ref{fig:Al-plate_systematics} and \ref{fig:80-20-geom_2_systematics}, respectively, as representative cases.

Similarly the systematic effect arising from the uncertainty in the electrical conductivity was evaluated by repeating the COMSOL calculations after shifting the electrical conductivity of Al6061-T6 by $\pm 8$\% and Al6063-T6 by $\pm 9$\%. The resulting change in the calculated magnetic-noise amplitude is shown in the bottom panel of Figs.~\ref{fig:Al-plate_systematics} and \ref{fig:80-20-geom_2_systematics}, respectively, as representative cases.

In all cases, the shifts in the calculated magnetic Johnson noise in the low frequency region agree with the expectation from, for example, the known expression for an infinite plane of thickness $t$ for low frequencies\cite{Varpula1984}
\begin{equation}
B_n (f=0) = \frac{\mu_0}{2} \sqrt{\frac{\sigma kT}{2\pi}\frac{t}{a(a+t)}},
\label{eq:zero_frequency_Johnson_noise}
\end{equation}
where $\mu_0$ is the vacuum permeability, $k$ is the Boltzmann constant, $T$ is the conductor temperature, $\sigma$ is its electrical conductivity, and $a$ is the distance between the observation point and the sample.  However, the shifts in the calculated magnetic Johnson noise exhibit complex frequency dependence for higher frequencies. This behavior is attributed to both the skin-depth effect and the self-inductance effect discussed below.   

Additional smaller experimental uncertainties include those from the temperature uncertainty and the uncertainty in the absolute SQUID calibration.

The temperature uncertainty results in uncertainties in the magnetic Johnson noise directly through the temperature factor in Eq.~(\ref{eq:F-Dtheorem}) as well as through the temperature dependence of the electrical conductivity. Therefore, in general, the temperature dependence of the magnetic Johnson noise is given by
\begin{equation}
    \frac{\Delta B_n}{B_n} \simeq \frac{1}{2}\frac{\Delta T}{T} + \frac{\partial \ln B_n}{\partial \ln \sigma}\frac{\Delta \sigma}{\sigma}.
    \label{eq:temperature_dependence}
\end{equation}
When the temperature changes from 300~K to 310~K for example ($\frac{\Delta T}{T} = 3.3$\%), the conductivity drops by 3 to 4\%.
At low frequencies, $\frac{\partial \ln B_n}{\partial \ln \sigma} = \frac{1}{2}$, as seen from Eq.~(\ref{eq:zero_frequency_Johnson_noise}). Therefore the first and second terms in Eq.~(\ref{eq:temperature_dependence}) can cancel each other almost exactly. 
However, this cancellation does not generally hold throughout the measured frequency range because the dependence of the magnetic Johnson noise on the electrical conductivity changes as seen in Figs.~\ref{fig:Al-plate_systematics} and \ref{fig:80-20-geom_2_systematics}. Using the dependence of the magnetic Johnson noise on the electrical conductivity shown in those figures, the shift in the magnetic Johnson noise due to temperature shift of $\pm 10$~K is estimated to be up to $\pm 2.4$\% at 3~kHz. 

The uncertainty in the absolute SQUID calibration is 2.0\%.\cite{Kim2026} These uncertainties are smaller than the ones included in the band shown in the bottom panel of Figs.~\ref{fig:Al_plate}, \ref{fig:80-20_geom_1}, \ref{fig:80-20_geom_2}, and \ref{fig:80-20_geom_3}.

Mesh convergence was evaluated by repeating the calculation using two progressively coarser meshes. The change in the calculated result decreased with increasing mesh density, indicating that the numerical calculation was approaching convergence. We took the difference between the results obtained with the baseline mesh and the next coarser mesh as the mesh-related uncertainty. This uncertainty was frequency dependent and varied among geometries and gradiometers, reaching approximately 2\% below 600~Hz and approximately 3\% over the full comparison range up to 3~kHz.

The uncertainty associated with mesh convergence does not represent a fundamental limitation of the approach. We deliberately performed the calculations on a laptop computer to demonstrate that the method is practical with modest computational resources. The numerical uncertainty can be reduced by further mesh refinement, at the cost of increased computation time and memory requirements. The computer used was a 2023 MacBook Pro with an Apple M2 Max processor and 64 GB of memory. It took about 1.5~h to compute one geometry with 4 gradiometers with a frequency sweep from 1 Hz to 7000 Hz with 10 points per decade. 

Between 1 and 3000 Hz, the broadband ratios ($B_{\rm meas}/B_{\rm calc}$) generally remain consistent with unity within this systematic uncertainty for all four geometries and all three operational gradiometers when the point-to-point scatter of 5\% due to the statistical uncertainty is considered. 
No sample geometry or conductivity parameters were adjusted to obtain agreement with the measured spectra.


Two clear changes in slope are apparent in the calculated and measured spectra. For the Al6061-T6 plate, the first occurs near 30~Hz, approximately where the skin depth becomes comparable to the 1.9-cm plate thickness. In addition to skin-depth effects, the self-inductance of the fluctuating current paths suppresses magnetic Johnson noise as their inductive reactance becomes comparable to their resistance. The characteristic frequencies for these two effects scale as \(f_{\mathrm{skin}}=1/(\pi\mu_0\sigma t^2)\) and \(f_{\mathrm{ind}}\sim1/(C\mu_0\sigma ta)\), respectively, where \(a\) is the distance to the conductor and \(C\) is a geometric factor of order unity.\cite{Lee2008} (Further discussion on the skin-depth and inductance effects is given below.) Because the plate thickness and the plate–gradiometer distance are comparable in our geometry, these characteristic frequencies are of the same order, and both effects contribute to the observed roll-off. 

For the 80/20 corner bracket, an initial change in slope occurs around 100 Hz, followed by a further, less pronounced change over approximately 300–500 Hz. Skin-depth effects and inductive screening due to the self-inductance of the fluctuating current paths both contribute to the frequency dependence. However, the bracket’s complex geometry precludes assigning a single thickness or current-loop size, so the two observed changes cannot be uniquely attributed to these mechanisms individually. The full finite-element calculation accounts for both effects and reproduces the observed spectral behavior. The additional steepening near 3 kHz results from the low-pass filter in the SQUID signal chain.

Further insight is gained by studying the frequency dependence of the change in magnetic Johnson noise when the geometry and the electrical conductivity are shifted, shown in Figs.~\ref{fig:Al-plate_systematics} and \ref{fig:80-20-geom_2_systematics}.

The increasing sensitivity to the gradiometer–sample distance at higher frequencies can be understood qualitatively from the skin-depth effect. At low frequencies, the detected magnetic Johnson noise receives contributions from regions throughout the sample thickness. When the skin depth becomes smaller than the thickness, these contributions become increasingly concentrated near the surface facing the gradiometers. The effective distance to the contributing region therefore decreases, so a fixed change in the gradiometer–sample separation represents a larger fractional distance change and can produce a larger relative change in the detected noise. The full finite-element calculation also accounts for inductive screening and the spatial response of the gradiometers.

The frequency dependence of the electrical conductivity-induced shifts can be understood qualitatively using the effective current-loop model of Lee and Romalis.\cite{Lee2008} In this model, the current-noise amplitude is given by
\begin{equation}
    \delta I = \frac{\sqrt{4kTR(f)}}{\left | R(f) + i2\pi f L\right |}.
\label{eq:current_noise}
\end{equation}
Here \(R(f) = \frac{2 \pi}{\sigma t_{\rm eff}}\), where $t_{\rm eff}$ is the effective thickness that includes the effects of the skin depth $\delta_{\rm skin}=1/\sqrt{\pi \mu_0 \sigma f}$, and \(L\) is the self-inductance of the current path. The numerator corresponds to the Johnson-Nyquist voltage noise and the denominator corresponds to the impedance of the circuit. For the self-inductance $L$, in the case of an infinite conductor plane or a large conductor plate as in our Al6061-T6 plate, we can assume a single effective annular loop of mean radius $\approx a$ and width $\approx a$,\cite{Lee2008} in which case $L=C \mu_0 a$, where $C$ is a constant of order unity. In the case of the 80/20 corner bracket, a more concrete estimate could be made since it forms a conductor loop. 

At low frequencies, resistance dominates the impedance and \(R\propto\sigma^{-1}\), giving \(B_n\propto\sqrt{\sigma}\). Increasing the electrical conductivity therefore increases the noise. As frequency increases, inductive reactance becomes important and reduces the sensitivity to electrical conductivity. When inductive reactance dominates, increasing the electrical conductivity instead decreases the noise because it reduces the resistive voltage-noise source. In the high-frequency limit where the skin depth is smaller than the sample thickness, \(R\propto\sigma^{-1/2}\), and the model gives \(B_n\propto\sigma^{-1/4}\). 

\begin{figure}[!t]
    \centering
    \includegraphics[width=1\linewidth]{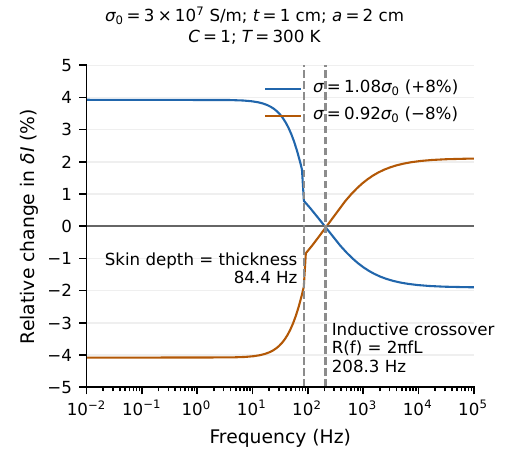}
    \caption{Numerical results from Eq.~(\ref{eq:current_noise}) on the shift in the Johnson noise current corresponding to a change in the electrical conductivity of $\pm 8$\%,  with $\sigma =3\times 10^7$~S/m, $t=1$~cm, $a=2$~cm, and $C=1$. }
    \label{fig:conductivity_change_model}
\end{figure}

Figure~\ref{fig:conductivity_change_model} shows numerical results from Eq.~(\ref{eq:current_noise}) on the shift in the magnetic Johnson noise corresponding to a change in the electrical conductivity of $\pm 8$\%, with $\sigma =3\times 10^7$~S/m, $t=1$~cm, $a=2$~cm, and $C=1$. The general features obtained agree well with what is shown in the bottom panel of Figs.~\ref{fig:Al-plate_systematics} and \ref{fig:80-20-geom_2_systematics}, while the detailed frequency dependence is determined by the sample geometry and gradiometer response. Note that the effect of the skin depth is implemented using a piecewise effective-thickness approximation, where the effective thickness is given by $t_{\rm eff} = \min(t,\delta_{\rm skin}(f,\sigma))$. The sharp bends near 84 Hz come from this replacement; a more complete skin-effect model would smooth them.

For the corner bracket, the electrical conductivity dependence exhibits additional structure near 1 kHz, particularly for gradiometers 2 and 4. This behavior is not captured by a single effective current-loop model and may reflect the changing contributions of current paths with different characteristic dimensions, weighted by the spatial response of each gradiometer.

In addition to the measured noise spectrum from the second-order gradiometer introduced above, Fig.~\ref{fig:4th_order_grad} shows the calculation using the F-D FEM method and the quadrature sum of the measured noise amplitudes of gradiometers 2 and 4. The latter represents the noise expected in their difference signal if the two signals were uncorrelated. The measured difference signal lies substantially below this quadrature sum, demonstrating positive correlation between the magnetic signals detected by the two gradiometers. The calculated noise amplitude agrees with the measured value within approximately 10\% below 100~Hz. Above this frequency, the measured amplitude increasingly exceeds the calculation, with a discrepancy of approximately 30–40\% near 1~kHz and above, excluding narrow spectral peaks. The individual gradiometers show substantially better agreement with the calculation. Because the second-order gradiometer measures their difference, its noise is particularly sensitive to spatial correlations and differences in the responses of the two detectors. The measured relative phase between the channels was consistent with zero, and changing to a coarser mesh resulted in a change of only approximately 2\% in the calculated noise amplitude. Therefore, neither the measured relative phase nor the observed mesh dependence accounts for the discrepancy. Other possible contributions to the discrepancy include differences between the actual and modeled coil geometries and electrical conductivities whose effects may be less apparent in the individual noise spectra. However, we did not investigate the origin of this discrepancy further.

Because this approach does not rely on geometry- or frequency-specific analytical approximations, its accuracy is limited primarily by the fidelity of the geometrical and material-property inputs and by the numerical accuracy of the finite-element calculation. The complex frequency dependence is particularly sensitive to both the geometry and the electrical conductivity. Therefore, in addition to calculating the expected magnetic Johnson noise during the design of an experiment, as demonstrated in Ref.~\onlinecite{Blatnik2026}, and interpreting experimental results, the method could in principle be used to determine the electrical conductivity of a material or constrain unknown geometrical parameters of a conducting structure. 

\section{\label{sec:summary}Conclusions}
We have experimentally validated the fluctuation–dissipation finite-element approach described in Ref.~\onlinecite{Phan2024} for calculating magnetic Johnson noise from conductors with complex geometries. The method reproduces the measured noise amplitudes and spectral shapes for multiple sensors and substantially different relative geometries. The combined-gradiometer comparison further demonstrates its versatility by testing spatial correlations between detector signals, with good agreement for low frequencies and an unresolved discrepancy at higher frequencies. The method is suitable for designing and interpreting precision magnetic measurements containing irregular conductive components. In addition, the method could also be used to measure the electrical conductivity of materials or to test or confirm geometries of a structure made of materials with known properties. 

\section{Acknowledgments}
This work was supported by the Laboratory Directed Research and Development program of Los Alamos National Laboratory Project No. LDRD20240078DR, the United States Department of Energy, Office of Science, Office of Nuclear Physics Grant No. DE-SC0025564, and the US National Science Foundation Grant No. NSF-2412368.

\section{Data Availability}
The data that support the findings of this study are available from the corresponding author upon reasonable request.

\bibliography{magnetic_johnson_noise}

\end{document}